\documentstyle[12pt]{article}
\begin{document}
\begin{flushright}
DF/IST-2.94
\end{flushright}
\vspace{.5cm}
\begin{center}
\huge An Exact Calculation of the Energy Density of Cosmological 
Gravitational Waves
\end{center}   
\vspace{3ex}    
\centerline{\Large L. E. Mendes\footnote{email: m1480@beta.ist.utl.pt} 
{\normalsize and } A. B. Henriques\footnote{email: alfredo@ptifm.bitnet}}  
%\vspace{.5ex} 
\begin{center} 
Dep. de F\'\i sica, Instituto Superior T\'ecnico, Av. Rovisco Pais,\\
1096 Lisboa Codex, Portugal
\end{center}
\vspace{2ex}
\centerline{\Large R. G. Moorhouse\footnote{email: 
gordonm@physics.gla.ac.uk}}
\begin{center}
Dept. of Physics and Astronomy, University of Glasgow,\\
Glasgow, G12 8QQ, Scotland
\end{center}

\vspace*{2cm}
 
\centerline{\bf \large  Abstract}
\noindent
In this paper we calculate the Bogoliubov coefficients and the 
energy density of the stochastic gravitational wave background for a 
universe that undergoes inflation followed by radiation domination and 
matter domination, using a formalism that gives the Bogoliubov coefficients 
as continous functions of time. By making a reasonable assumption for the 
equation of state during reheating, we obtain in a natural way the expected 
high frequency cutoff in the spectral energy density.

\vspace{2ex}
\newpage
\baselineskip 0.5cm  
\section{Introduction} 
\hspace{\parindent}
In the usual framework for the calculation of the stochastic gravitational 
wave background spectrum \cite{ah} it is assumed that the transition between 
two consecutive epochs in the evolution of the universe is sudden, 
requiring only the continuity of the metric and its first derivative at 
the transition point (when the unperturbed space-time manifold is the 
Friedmann-Lema\^{\i}tre-Robertson-Walker (FLRW) model, this reduces to the 
continuity of the scale factor and its first derivative). Matching 
the wave function for the perturbation and its first derivative at 
both sides of the transition, we obtain the Bogoliubov coefficients    
relating the initial and the final states. This sudden transition 
approximation gives correct results, except for very large frequencies  
where it gives an overproduction of gravitons that causes the energy 
density to diverge \cite{allen}. This overproduction can be traced back to the fact 
that in this approximation, the second derivative of the scale factor and  
therefore the scalar curvature is discontinuous \cite{ford}. To solve this 
problem, a 
cutoff is introduced for frequencies $f$ larger than the rate of expansion 
$f_t$ at 
the time the transition occurs, by invoking general adiabatic considerations 
\cite{parker}, \cite{bd}. Loosely speaking, the modes with $f \gg f_t$ 
will not notice that the universe is expanding and hence 
will be exponentially suppressed; this suppression is well simulated  
by the cutoff. But for frequencies near $f_t$ the problem cannot be  
correctly solved by the use of a sharp cutoff.

Contrary to what happens with density perturbations where the high frequency 
part of the spectrum can be easily distorted, 
the high frequency part of the gravitational wave spectrum  
will remain unchanged during the evolution of the universe
due to the fact that gravitational waves decouple very early. 
Thus, 
if our goal is the comparison of our spectra either with data from 
direct observations \cite{liddle} or Cosmic Microwave Background Radiation 
measurements (CMBR) \cite{cobe} \cite{white-sour}, 
more attention should be given to the high
frequency part of the gravitational wave background spectrum.

In this paper we calculate the exact Bogoliubov coefficients 
for the production of the gravitational wave background using a method 
developed in \cite{us}. This method does not rely on the sudden transition  
approximation and can be used to obtain the Bogoliubov coefficients for 
the high frequency modes without any approximation. 
Making a reasonable assumption about the equation of state during reheating 
and using a scale factor that interpolates smoothly between a radiation 
dominated universe and a matter dominated universe we obtain in a natural way 
the high frequency cutoff.

The outline of this paper is as follows: in section II we summarize our 
method for the calculation of the Bogoliubov coefficients; in section III we 
calculate the exact Bogoliubov coefficients for a universe that undergoes 
inflation, reheats and evolves through a radiation dominated epoch followed 
by a matter dominated epoch; we end the paper with our conclusions in 
section IV.

\section{The method}
\hspace{\parindent}
For simplicity we assume that the unperturbed space-time manifold is the flat 
FLRW model. The perturbed metric can be expressed in the synchronous gauge as
\begin{equation}
ds^2 = a^2(\tau) \left[ - d\tau^2 + \left( \delta_{ij} + 
h_{ij}(\tau,{\bf x})\right) \, dx^i dx^j \, \right]
\label{flrw}
\end{equation}
where $\tau$ is the conformal time, $a(\tau)$ is the scale factor and 
$h_{ij}(\tau,{\bf x})$ 
is the perturbation (latin indices run from $1$ to $3$). Expanding $h_{ij}$ 
in plane waves we get 
\begin{eqnarray}
\lefteqn{h_{ij}(\tau,{\bf x}) = \sqrt{8 \pi G} \sum_{p=1}^{2} \int 
\frac{d^3 k}{(2 \pi)^{3/2} a(\tau) \sqrt{2 k}} \times }  \nonumber \\
& & \left[ a_{p}({\bf k},\tau) \varepsilon_{ij}({\bf k},p) 
e^{i {\bf k} \cdot {\bf x}} \xi(k,\tau) \; + \; herm. \; conj. \right]
\end{eqnarray}
where ${\bf x}$ is the spatial coordinates three-vector, ${\bf k}$ is the 
comoving wave-number three-vector, $k = |{\bf k}|=\frac{2 \pi a}{\lambda}= 
\omega a$, $p$ runs over the two possible polarizations of the gravitational 
waves, $\varepsilon_{ij}({\bf k},p)$ is the polarization tensor,  
$a_{p}({\bf k},\tau)$ the annihilation operator and $\xi(k,\tau)$ the mode 
function 
for the gravitational waves obeying the equation 
\begin{equation}
\xi'' + (k^2 - \frac{a''}{a}) \xi = 0
\end{equation}
where the $'$ denotes the derivative with respect to the conformal time.
The annihilation and creation operators can be expressed in terms of the 
initial time annihilation and creation operators $A_{p}({\bf k})$ and 
$A^{\dagger}_{p}({\bf k})$  through the Bogoliubov coefficients 
$\alpha(k,\tau)$ and $\beta(k,\tau)$
\begin{equation}
a({\bf k},\tau) = \alpha(k,\tau) A({\bf k}) + 
\beta^{\ast}(k,\tau) A^{\dagger}({\bf k}) 
\end{equation}
(we have dropped the polarization subscripts for clarity) where the 
$\alpha$ and $\beta$ must satisfy
\begin{equation}
| \alpha|^2 - | \beta|^2 = 1   \; .
\end{equation}

In ref. \cite{us} it has been shown that the Bogoliubov coefficients obey the set 
of coupled ODE's
\begin{eqnarray}
\alpha'(\tau) & = & \; \; \frac{i}{2 k} \left( \alpha(\tau) + \, \beta(\tau) 
e^{\, 2 i k (\tau - \tau_0)}\right)  \frac{a''(\tau)}{a(\tau)} 
\label{bogode1} \\
\beta'(\tau) & = & - \frac{i}{2 k} \left( \beta(\tau) + \alpha(\tau) 
e^{- 2 i k (\tau - \tau_0)} \right)  \frac{a''(\tau)}{a(\tau)} \label{eq2}     
\label{bogode2}
\end{eqnarray}
where $\tau_0$ is an arbitrary constant. These equations are equivalent 
to Parker's integral equations \cite{parker} for $\alpha$ and $\beta$ 
provided we make for his $W(k,\tau)$ the ansatz $W(k,\tau)=k$. Substituting  
\begin{eqnarray}
\alpha & = & \frac{1}{2} (X + Y) e^{\, i k (\tau - \tau_0)} \label{xydef1} \\
\beta & = & \frac{1}{2} (X - Y) e^{- i k (\tau - \tau_0)} 
\label{xydef2}
\end{eqnarray}
with $X \equiv X(k,\tau)$ and $Y \equiv Y(k,\tau)$ our system takes the form
\begin{eqnarray}
X'' + (k^2 - \frac{a''}{a}) X & = & 0 \label{xeqdif}  \label{xeq} \\
Y & = & \frac{i}{k} X'   \;  .
\label{yeq}
\end{eqnarray}
When the scale factor has the power-law form, the solution of equation 
(\ref{xeqdif}) can be expressed in terms of the Hankel functions 
\cite{marcio}. Once we have 
solved equation (\ref{xeqdif}) for the entire period under investigation, 
$\alpha$ and $\beta$ can be easily 
calculated from equations (\ref{yeq}), (\ref{xydef1}) and (\ref{xydef2}). 
To obtain the final Bogoliubov coefficients we must now perform one more 
Bogoliubov transformation to the mode functions appropriate to the 
present state of the universe as explained in ref. \cite{us}. Denoting the 
readily calculable Bogoliubov coefficients corresponding to the final 
transformation by $\alpha_{ft}$ and $\beta_{ft}$, the final Bogoliubov 
coefficients $\alpha_{F}$ and $\beta_{F}$ will then be given by 
\begin{eqnarray}
\alpha_{F} & = & \alpha \, \alpha_{ft} + \beta \, \beta^{\ast}_{ft} 
\label{finalalpha}\\
\beta_{F} & = & \beta \, \alpha^{\ast}_{ft} + \alpha \, \beta_{ft}  \; .
\label{finalbeta}
\end{eqnarray}

With the formalism described above no sudden transition approximation is 
needed and the high frequency part of the graviton spectrum can be 
obtained exactly, provided we can make some reasonable assumption about the 
behaviour of the scale factor during the transitions between two consecutive 
epochs.

\section{The Bogoliubov coefficients}
\hspace{\parindent}
We will consider a Universe that undergoes inflation, reheating and enters 
an epoch where dynamics is governed by a mixture of radiation and dust. The 
scale factor has the form 
\begin{equation}
a(\tau) = \left\{ 
\begin{array}{ll}
\frac{1}{H (\tau_1 - \tau)} & \tau \leq \tau_I \\
a_R(\tau) & \tau_I < \tau \leq \tau_R \\
a_{eq} \left( \tau^{2} + 2 \tau \right) & \tau > \tau_R
\end{array}
\right.  
\label{scaf}
\end{equation}
where $a_{R}(\tau)$ is an yet undetermined function of $\tau$ that will 
smoothly interpolate between inflation and the next epoch, $a_{eq}$ is the 
scale factor at the time of equality between the densities of radiation and 
dust $\tau_{eq}$ and $H$ (the Hubble constant during inflation) and 
$\tau_{1}$ will be determined later. The solution is normalized in such a 
way that $\tau_{eq} = \sqrt{2} - 1$. The scale factor for 
$\tau > \tau_{R}$ is a solution of the Friedmann equation for a universe 
filled with a non-interacting mixture of radiation and dust. This assumption 
provides a 
smooth transition from a radiation dominated universe to a dust dominated 
universe and is the simplest possible improvement over the usual treatment,  
certainly giving a better approximation near $\tau_{eq}$; 
for $\tau \ll \tau_{eq}$ the scale factor reduces to the one for a radiation 
dominated universe 
$a \propto \tau$ and for $\tau \gg \tau_{eq}$ we recover the scale factor for 
a dust dominated universe $a \propto \tau^2$. 

We still have to do some assumptions about the scale factor during reheating. 
A reasonable way to procede is to assume an independent equation of state 
for each of 
the components of the fluid filling the universe and 
integrating the Friedmann and energy density conservation equations to obtain 
$a_{R}(\tau)$. A simple but not unique choice for the equation of state is
\begin{equation}
p_{i}(\tau) = \lambda_{i}(\tau) \, \rho_{i}(\tau) \;\;\;\;\;\;\; (i=1,2)
\label{eqstate}
\end{equation}
where the $\lambda_{i}$ must satisfy the limiting values
\begin{equation}
\begin{array}{cr}
\lambda_1(\tau_I) = \lambda_2(\tau_I) = -1 & \\
\lambda_1(\tau_R) = \frac{1}{3} &  \\
\lambda_2(\tau_R) = 0        & \;\; ,
\end{array} 
\label{lambdacond}
\end{equation}                       
as appropriate to the inflationary and radiation$+$dust stages respectively.
Of course, our choice of equation of state does not intend to be more than  
illustrative of the method we are using.
From the present values of $\rho_{rad}$ and $\rho_{dust}$ we obtain 
$\rho_{i}(\tau_{R})$
\begin{eqnarray}
\rho_1(\tau_R) = \rho_{rad}(\tau_R) = \rho_{rad}(\tau_p) 
\left( \frac{a(\tau_p)}{a(\tau_R)} \right)^4  \label{rho1ini} \\
\rho_2(\tau_R) = \rho_{dust}(\tau_R) = \rho_{dust}(\tau_p) 
\left( \frac{a(\tau_p)}{a(\tau_R)} \right)^3 
\label{rho2ini}
\end{eqnarray}
where $\tau_{p}$ is the present conformal time.

Denoting by $f(\tau)$ the function
\begin{equation}
f(\tau) = \left\{ \begin{array}{ll} 
0 & \tau \leq 0 \\
\exp (- c_1 / \tau) & \tau > 0
\end{array}
\right.
\label{eqstatef}
\end{equation}
we choose $\lambda_{i}$ as 
\begin{eqnarray}
\lambda_1 & = & \frac{1}{3} 
\frac{f(\tau - \tau_I)}{f(\tau - \tau_I) + f(\tau_R - \tau)} -
\frac{f(\tau_R - \tau)}{f(\tau - \tau_I) + f(\tau_R - \tau)}
\label{lambda1} \\
\lambda_2 & = & 
- \frac{f(\tau_R - \tau)}{f(\tau - \tau_I) + f(\tau_R - \tau)}
\label{lambda2}
\end{eqnarray}
where the constant $c_1$, can be used to control the rate 
at which the transition occurs. It can be easily seen that these $\lambda$'s 
obey conditions (\ref{lambdacond}). As fig. 1 shows, these are very 
smooth functions.

We can finally write down the equations that determine $a_R$, $\rho_1$ and 
$\rho_2$ 
\begin{eqnarray}
\left( \frac{a'}{a^2} \right)^2 & = & 
\frac{8 \pi G}{3 c^2} \left( \rho_1 + \rho_2 \right) \;\;\;\;\;\;\;\;\;\;\;\;
\label{friedmann} \\
\rho'_{i} & = & - 3 \frac{a'}{a} \rho_i \left( 1 + \lambda_i \right)  
\;\;\;\;\; (i=1,2)
\label{rhoi}
\end{eqnarray}
with the $\lambda_i$ given by (\ref{lambda1}) and (\ref{lambda2}).
These equations can be easily solved numerically. If we want our model to 
be consistent with the present day observed universe then we must 
evolve back in time the present condition of the universe to obtain the 
state of the universe at $\tau_R$ and then we integrate equations 
(\ref{friedmann}) and (\ref{rhoi}) backwards in time up to $\tau_I$. 
This procedure will give us a scale factor that is compatible with the 
observed Universe.
Once the numerical integration is done we can easily find $H$ and 
$\tau_1$ appearing in eq. (\ref{scaf})
\begin{equation}
H = \frac{a^2_R(\tau_I)}{a'_R(\tau_I)} \, ; \;\;\;\; 
\tau_1 = \frac{a_R(\tau_I)}{a'_R(\tau_I)} + \tau_I
\label{ht1}
\end{equation}
In our calculations we used for the redshifts at $\tau_{eq}$, $\tau_R$ 
and $\tau_I$ the values $1 + Z_{eq} = 2.38 \times 10^4$,
$1 + Z_{R} = 3 \times 10^{28}$ and $1 + Z_{I} = 1.6 \times 10^{29}$. 

Knowing $a(\tau)$ for the entire period under consideration,
we are now prepared to calculate the Bogoliubov coefficients.
The initial conditions for $\alpha$ and $\beta$ are
\begin{equation}
|\alpha(- \infty)| = 1 \, ; \;\;\;\; |\beta(- \infty)| = 0
\label{incondab}
\end{equation}
which translate for the $X$ and $Y$ variables into 
\begin{equation}
|X(- \infty)| =| Y(- \infty)| = 1
\label{incondxy}
\end{equation}
Equation (\ref{xeq}) can be integrated analyticaly up to $\tau = \tau_I$. 
The solution that satisifes (\ref{incondxy}) gives at $\tau_I$
\begin{eqnarray}
X(\tau_I) & = & e^{- i k (\tau_I - \tau_1)} 
\left( 1 + \frac{i}{k (\tau_1 - \tau_I)} \right)
\label{xyti1}  \\
Y(\tau_I) & = & e^{- i k (\tau_I - \tau_1)} 
\left( 1 + \frac{i}{k (\tau_1 - \tau_I)} - 
\frac{1}{k^2 (\tau_1 - \tau_I)^2} \right)
\label{xyti2}
\end{eqnarray}
At this point we must procede with the integration numerically using 
(\ref{xyti1}) and (\ref{xyti2}) as initial conditions. After recovering 
$\alpha$ and $\beta$ from (\ref{xydef1}) and (\ref{xydef2}) we must 
perform the final Bogoliubov transformation  (\ref{finalalpha}) and 
(\ref{finalbeta}). To do this, we notice first that in spite of our final 
state being one of mixed radiation and dust, as $\tau_p \gg \tau_{eq}$ 
the contribution of the radiation component is negligible (we have seen 
at the beginning of section III that for $\tau \gg \tau_{eq}$ we recover 
the scale factor for a matter dominated universe $a \propto \tau^2$) 
and we can consider that the final state in our model is one of matter 
domination. As explained fully in ref. \cite{us} the final Bogoliubov 
coefficients must be those appropriate to the present day 
matter era mode functions and this requires a final 
Bogoliubov transformation to the matter era mode functions with the 
corresponding   
Bogoliubov coefficients $\alpha_{ft}$ and $\beta_{ft}$ in 
equations (\ref{finalalpha}) and (\ref{finalbeta}) given by 
\cite{ah}, \cite{allen}, 
\cite{us}
\begin{eqnarray}
\alpha_{ft} & = & e^{i k \tau_0}
\left( 1 + \frac{i}{k \tau_p} - \frac{1}{2 (k \tau_p)^2} \right)
\label{alphaft} \\
\beta_{ft} & = & - \frac{e^{- i k (2 \tau_p - \tau_0)}}{2 (k \tau_p)^2} \;\; .
\label{betaft}
\end{eqnarray}

Figures 2 shows our final results expressed as the energy
density $P(\omega) = dE/d\omega = \hbar \omega^3 |\beta|^2 / \pi^2 c^3$. 
It is clear from this figure that with the 
approach developed in this paper it is possible to obtain, in a natural 
way the high frequency cutoff.

For the sake of comparison we also calculate the Bogoliubov coefficients 
in the sudden transition approximation for a universe that   
undergoes inflation, radiation domination and matter domination.
Assuming that in this case the transition between inflation and radiation
occurs at $\tau_{ir}$ and denoting the Bogoliubov coefficients associated 
with the transition from inflation to radiation domination and from radiation 
to matter domination respectively with the subscripts $r$ and $d$ we have  
\begin{eqnarray}
\alpha_{r} & = & e^{2 i k \tau_I'} \left( 1 + \frac{i}{k \tau_I'} 
- \frac{1}{2 (k \tau_I')^2} \right) \label{alfar} \\
\beta_r & = & \frac{1}{2 (k \tau_I')^2}  \label{betar} \\
\alpha_{d} & = & e^{i k \tau_{eq}} \left( 1 + \frac{i}{2 k \tau_{eq}} 
- \frac{1}{8 (k \tau_{eq})^2} \right) \label{alfad} \\
\beta_d & = & - e^{- 3 i k \tau_{eq}} \frac{1}{8 (k \tau_{eq})^2} 
\label{betad}
\end{eqnarray}
The final Bogoliubov coefficients are then given by \cite{ah},\cite{allen}
\begin{equation}
\alpha = \left\{ \begin{array}{ll} 
\alpha_{r} & k_{min} < k \leq k_{r} \\
\alpha_{r} \, \alpha_{d} + \beta_{r} \, \beta^{\ast}_{d} &
k_{r} < k \leq k_{cut} 
\end{array}
\right.
\end{equation}
and
\begin{equation}
\beta= \left\{ \begin{array}{ll} 
\beta_{r} & k_{min} < k \leq k_{r} \\  
\beta_{r} \, \alpha^{\ast}_{d} + \alpha_{r} \, \beta_{d} & 
k_{r} < k \leq k_{cut}
\end{array}
\right.
\end{equation}
with 
\begin{eqnarray}
k_{min} & = & 2 \, \pi \, a(\tau_p) \, H(\tau_p) \\
k_{r} & = & 2 \, \pi \, a(\tau_{eq}) \, H(\tau_{eq}) 
\end{eqnarray}
To fix the value of $k_{cut}$ we notice that, acording to general adiabatic  
arguments, the characteristic frequency for the cutoff is given by 
$\omega_{cut} = 2 \, \pi \, \Delta t^{-1}$ where $\Delta t$ is the (physical)
time scale for the transition, usualy taken to be $H(\tau_{ir})^{-1}$ 
\cite{marcio}, \cite{varun}. However, as these authors point, the exact 
value of $\Delta t$ depends on the details of the transition. In our model  
$\Delta\tau$ is known and we have $\Delta t \approx a(\tau_{ir}) \Delta\tau$ 
which gives
\begin{equation}
k_{cut} = \frac{2 \pi}{\Delta\tau}  \;\;\;\;\; .
\label{kcut}
\end{equation}

At this point we still have to determine $\tau_{ir}$. Because in our model 
we have introduced one intermediate epoch between inflation and radiation 
there is, when comparing the two models, an ambiguity in the choice of $\tau_{ir}$.
To overcome this ambiguity we choose $\tau_{ir}$ in such a way that the
number of gravitons at the beginning of the radiation dominated era 
is the same in the two models. 

Figure 3 shows the comparison of the results obtained with our method
and the one using the sudden transition approximation. 
We can see in Fig. 3a that the method discussed in this paper 
predicts a larger $P$ for the low frequency gravitons than is the case with 
the usual method of calculation. This can be easily 
explained by the fact that during the radiation dominated epoch, and if we
assume only radiation to be present, there is 
no graviton production due to the well known conformal invariance of the 
equations \cite{parker}, \cite{grishuk}, \cite{gun}. 
In our model the invariance is 
broken because we also have dust mixed with the radiation and we thus have an 
additional contribution to $P$. No matter 
how small the contribution of the dust component becomes, the conformal   
invariance will always be broken, as can be seen by the fact that 
$a'' = a_{eq} \neq 0$ in conformal time. We should also notice (fig. 3b) 
that the point where the high frequency suppression starts to take 
place, depends 
on the time it takes for the transition between inflation and radiation 
domination, as expected from adiabatic arguments \cite{parker} \cite{bd}.  
The dots in Fig. 3b represent the energy density obtained with the sudden 
transition approximation and $k_{cut}$ given by (\ref{kcut}) with 
$\Delta\tau=6. \times 10^{-26}$, corresponding to the width of the 
transition in our model with $c_1=11.$. 
Comparing the two curves, we can see that, for frequencies near the 
suppression, 
the energy density 
obtained by our exact calculation is one order of magnitude bellow the 
result obtained with the sudden transition approximation.  

\section{Conclusions}
\hspace{\parindent}
We have shown in this paper how the high frequency cutoff in the 
energy density of the stochastic gravitational wave background could be 
obtained in a natural way. Although our results are not surprising, the 
calculation done in this paper is exact, even in the high frequency region   
of the spectrum. We showed that for frequncies slightly below the high 
frequency cutoff, the sudden transition approximation predicts an 
overproduction of gravitational waves.
Differences to previous calculations were also  
obtained in 
the region of frequencies corresponding to the transition between the 
radiation dominated and the matter dominated epochs. 

\vspace{1cm}
\noindent
After this work was completed we came across a preprint by Koranda and 
Allen \cite{ka} where the epoch following inflation is also treated as 
a mixture of radiation and dust.

\vspace{2cm}
\noindent 
{\Large \bf Acknowledgements} \\

\vspace{.5cm}
\noindent 
The authors acknowledge the receipt of a NATO Grant for Colaborative 
Research CRG 9290129.
L.E.M. acknowledges the support of a JNICT grant under contract 
BD/2125/92/RM.

\newpage
\noindent
{ \Large \bf FIGURE CAPTIONS } \\

\vspace{.7cm}
\noindent
{\large \bf Figure 1} The $\lambda$'s given by equations (\ref{lambda1}) 
(a)and (\ref{lambda2}) (b) as functions of the 
conformal time $\tau$ during the 
transition from inflation to the radiation+dust era. We can see that the 
width of the transition can be controlled with the parameter $c_{1}$. 
Larger values of $c_{1}$ correspond to faster transitions.
We used $\tau_{I}=3.9585 \times 10^{-27}$ and 
$\tau_{R}=3.9585 \times 10^{-25}$ in all the figures.

\vspace{.4cm}
\noindent
{\large \bf Figure 2}
The energy density of the gravitational waves produced during the expansion 
of the universe obtained with the method developed in this paper. The steeper 
part of the curve ($f < 10^{-16} s^{-1}$) corresponds to gravitons 
produced during the  
radiation$+$dust epoch and the region with constant slope is due to graviton 
production during inflation. The sharp suppression is due to the finite width 
of the transition from inflation to radiation domination. This figure was obtained with $c_{1}=11$.

\vspace{.4cm}
\noindent
{\large \bf Figure 3}
For low frequencies, ($f < 10^{-16} s^{-1}$) the contribution for $P$ of the 
gravitons produced during the radiation$+$dust epoch is larger in our model 
than in the sudden transition approximation. Due to the presence of a small 
contribution of dust in the radiation dominated era, the conformal invariance  
of the equations is broken giving rise to more graviton production (a). 
As it was expected, the point where the high frequency suppression appears 
depends on the time it takes for the transition from inflation to radiation
domination. In our model the width of the transition can be varied with 
$c_{1}$ (b). The dots represent the results obtained with the sudden 
transition approximation; we introduced a high frequency 
cutoff $k_{cut}=2 \pi/\Delta\tau$ with $\Delta\tau = 6. \times 10^{-26}$ 
corresponding to the width of the transition for the case 
$c_1=11.$ (See Fig. 1).

\end{document}